\documentclass[pra,aps,showpacs,showkeys,twocolumn]{revtex4}
\usepackage{graphicx}
\usepackage{color}
\begin{document}
\title{\bf Chaplygin Gas Ho\v{r}ava-Lifshitz Quantum Cosmology}
\author{Hossein Ardehali}
\author{Pouria Pedram}
\email[Electronic address: ]{p.pedram@srbiau.ir}\affiliation{Department
of Physics, Science and Research Branch, Islamic Azad University,
Tehran, Iran}

\date{\today}

\begin{abstract}
In this paper, we study the Chaplygin gas Ho\v{r}ava-Lifshitz
quantum cosmology. Using Schutz formalism and Arnowitt-Deser-Misner
decomposition, we obtain the corresponding
Schr\"{o}dinger-Wheeler-DeWitt equation. We obtain exact classical
and quantum mechanical solutions and construct wave packets to study
the time evolution of the expectation value of the scale factor for
two cases. We show that unlike classical solutions and upon choosing
appropriate initial conditions, the expectation value of the scale
factor never tends to the singular point which exhibits the
singularity-free behavior of the solutions in the quantum domain.
\end{abstract}


 \maketitle

\section{Introduction}
In recent years, the results from supernova Ia have shown that the
expansion of our Universe is accelerating unlike the
Friedmann-Robertson-Walker (FRW) cosmological models with
nonrelativistic matter and radiation \cite{C1,C2,C3}. Also, the
cosmic background radiation data imply that our Universe is in
positively accelerated state \cite{C4,C5,C6}. The cosmological
constant $\Lambda$ as the usual vacuum energy may be responsible for
this accelerating evolution of the Universe with the negative
pressure. Unfortunately, the measured value of the cosmological
constant is 120 orders of magnitude smaller than its theoretical
predicted value \cite{C7,C8}.

On the other hand, thr Chaplygin gas, as a perfect fluid which
behaves like a pressureless fluid at early times and a cosmological
constant at later times, can be a candidate for the dark energy
\cite{ch_gas2,Benaoum,ZXL,visc}. The generalized Chaplygin gas with
negative pressure is described by an exotic equation of state,
\begin{equation}\label{eq1}
 p=-\frac{A}{\rho^\alpha},
\end{equation}
where $A$ is a positive parameter, $p$ is the pressure, $\rho$ is
the energy density, and $\alpha$ is a positive parameter so that
$0<\alpha\leq1$. In the standard model of Chaplygin gas, we set
$\alpha=1$ \cite{ch_gas4}. Recently, various studies have been done
in the literature, such as cosmology with the Chaplygin gas to
explain the transition from a dust-dominated Universe to the
accelerating expansion stage \cite{ch_gas1,ch_gas4}, modified
generalized Chaplygin gas \cite{MCG1,MCG2,MCG3}, modified cosmic
Chaplygin gas \cite{MCCG1,MCCG2}, and phenomenological relations
between brane-world scenarios and FRW minisuperspace cosmologies in
the presence of the generalized Chaplygin gas \cite{BLM}.

The idea of an early Chaplygin gas phase in the Universe was first
suggested in Refs.~\cite{BLM,L20,L21} and later extended in
Refs.~\cite{L22,L23,L24}. It is interesting to note that the
generalized Chaplygin gas has a connection with string theory. It
can be effectively obtained in the light cone parametrization from
the Nambu-Goto action for a $d$-brane in a ($d+1$,1)-dimensional
spacetime \cite{J25}. This motivates the application of the
generalized Chaplygin gas model at early time in the evolution of
the Universe. In particular, the Chaplygin inflation is addressed in
Ref.~\cite{L20} and the tachyon-Chaplygin inflationary Universe is
studied in Ref.~\cite{J27}.

In fact, Chaplygin spacetime models are related to a generalized
Born-Infeld action with a complex scalar field which corresponds to
a perturbed $d$-brane in a ($d + 1$,1)-dimensional spacetime
\cite{ch_gas2,ch_gas1,x31}. A generalized Born-Infeld phantom
inspired generalized Chaplygin gas model is presented in
Ref.~\cite{x30}. This opens a window to investigate brane-world
physics based on a phenomenological point of view. Brane-world
scenarios could explain various cosmological effects such as the
inflation based on the WMAP data \cite{C5}, the origin of our
Universe, and other fundamental issues such as the hierarchy problem
\cite{x34}. The quantum cosmological creation of brane-world models
is also addressed in Refs.~\cite{x37,x42,x43,x44,x45,x46,x47,x48}.
This allows us to use the similarities between the quantum cosmology
of a brane-world model and the Chaplygin gas quantum cosmology.

Recently, a single-field inflation scenario is presented in
Ref.~\cite{JCAP} where the properties of the inflaton field
areequivalent to a generalized Chaplygin gas. Based on the
measurements released by the Planck data and the WMAP large-angle
polarization, it is found that the $\alpha$ parameter is given by
$\alpha = 0.2578 \pm 0.0009$ \cite{JCAP,alpha,comment}. In
Ref.~\cite{Lopez2013}, a slow-roll inflationary scenario is modeled
by a generalized Chaplygin gas that can interpolate between a
network of frustrated topological defects and a de Sitter-like or a
power-law inflationary era.

After the Universe reached the dust-dominated stage, the
exponentially expanding evolution of the Universe represents a
quantum mechanical transition with some remnant component of the
original wave function of the Universe \cite{BLM}. In fact, there is
a quantum mechanical background behind the classically observed
Universe on large scales during the dust dominance epoch which is
based on the cosmological influence of a rapidly oscillating wave
function with a small amplitude. This wave function remnant can be
considered as a robust component with respect to decoherence
processes in the early times \cite{x49,x50,x51,x52}.

In 1970, Schutz introduced a velocity potential representation for
the four-velocity of a perfect fluid in general relativity
\cite{schutz1}. The equations of hydrodynamics for a perfect fluid
are expressed in terms of six scalar potentials $\mu$, $\phi$,
$\bar{\alpha}$, $\beta$, $\theta$, and $S$ so that
\begin{equation}\label{eq2}
        u_\nu=\frac{1}{\mu}(\partial_\nu\phi
        +\bar{\alpha}\,\partial_\nu\beta+\theta\,\partial_\nu S),
\end{equation}
where each of these potentials has their own equation of motion.
Indeed, these equations are equivalent to those based on divergence
of the stress-energy tensor \cite{schutz1}. Here, $\mu$ is the
specific enthalpy and $S$ is the specific entropy of the fluid. The
potentials $\bar{\alpha}$ and $\beta$ are related with rotations
and, hence, they are absent in the FRW Universe due to the symmetry
of the model. The potentials $\phi$ and $\theta$ do not have clear
physical meaning. Moreover, the four-velocity obeys the usual
normalization, namely, $u^\nu u_\nu=-1$.

The velocity potential version of the perfect fluid is based on the
variational principle with the following Lagrangian density
\begin{equation}\label{eq3}
        \mathcal{L}=\sqrt{-g}\,(R+16\pi\,p),
\end{equation}
where $R$ is the Ricci curvature scalar in four dimensions, $p$ is
the pressure of the perfect fluid, and $g$ is determinate of the
four-metric. Variation of this action with respect to the metric
gives rise to the Einstein field equations, and the variation with
respect to each of the velocity potentials yields the equations of
motion for the potentials \cite{schutz1,schutz2}. Also, in this
framework, the Arnowitt-Deser-Misner (ADM) formalism can be used to
decompose the spacetime metric $g_{\mu\nu}(t,\mathbf{x})$ in terms
of the three-dimensional metric $\gamma_{ij}(t,\mathbf{x})$, shift
vector $N^i(t,\mathbf{x})$, and the lapse function $N(t,\mathbf{x})$
to obtain the Hamiltonian and Einstein equations in three dimensions
\cite{ADM,schutz2}.

A new theory of gravity presented by Ho\v{r}ava is based on the
asymmetry scaling of space $\textbf{x}$ and time $t$, where it is
characterized by a scaling parameter $b$ and the dynamical critical
exponent $z$  as \cite{Horava1,Horava2,Horava3,Horava4}
\begin{equation}\label{eq4}
        \textbf{x}\rightarrow b\,\textbf{x},\qquad t\rightarrow b^z\,t.
\end{equation}
The resulting theory, the so-called Ho\v{r}ava-Lifshitz (HL)
gravity, is proved to be power-countable renormalizable. This theory
is based on the assumption that higher spatial-derivative correction
terms such as different powers of the spatial curvature and its
derivatives could be added to the standard Einstein-Hilbert action.
This result leads to improvement in the UV behavior of the graviton
propagator, but the Lorentz invariance as a fundamental symmetry of
theory is broken \cite{HL1}.

Ho\v{r}ava made an important assumption about the lapse function
which simplified the HL gravitational action, the so-called
``projectability condition,'' as $N\equiv N(t)$. Projectable
theories lead to a unique integrated Hamiltonian constraint, which
also cause great complications when they are compared with general
relativity. On the other hand, as in general relativity,
nonprojectable theories give rise to a local Hamiltonian constraint
\cite{nonprojHL}. However, since FRW spacetime is homogeneous and
isotropic, the spatial integral can be dropped from the integrated
Hamiltonian constraint \cite{HL2,HL3} which results in a true local
constraint even for the projectable case. Thus, in our study it is
sufficient to consider the FRW-HL projectable theory as the starting
point. Another assumption that is introduced by Ho\v{r}ava is the
principle of detailed balance. Based on this condition, the
potential in the gravitation action is originated from the gradient
flow generated by a three-dimensional action and reduces the number
of independent coupling constants. Notice that it has been recently
found that the detailed balance condition can also be relaxed
\cite{HL4,HL5,HL6,HL7}.

When the spacetime is asymmetric (\ref{eq4}), the dimensions of space and time are different, namely,
\begin{equation}\label{eq5}
        [\textbf{x}]=[\kappa]^{-1},\qquad[t]=[\kappa]^{-z},
\end{equation}
where $\kappa$ is a placeholder symbol with the dimensions of
momentum. For general values of $z$, the classical scaling
dimensions of the fields are given by
\begin{eqnarray}\label{eq5_2}
        [g_{ij}]=[N]=1,\qquad[N^i]=[\kappa]^{z-1},\qquad[ds]=[\kappa]^{-1},
\end{eqnarray}
and therefore $[dV_{d+1}]=[\kappa]^{-z-d}$ \cite{Lv}. Throughout the
paper, we take $z=3$.

The gravitational action of the HL model consists of the kinetic
part $S_K$ and the potential part $S_V$ in the form $S_{HL}=S_K +
S_V$. The kinetic part comes from the Einstein-Hilbert action and in
terms of the ADM variables takes the form
\begin{equation}\label{eq6}
        S_K=\int d^3xdt\,N\sqrt{\gamma}(K_{ij}K^{ij}-\lambda K^2),
\end{equation}
where $\lambda$ is coupling constant and $K_{ij}$ is the exterior
curvature tensor,
\begin{equation}
K_{ij}=\frac{1}{2N}\left(\dot{\gamma}_{ij}-\nabla_iN_j-\nabla_jN_i\right),\qquad
i,j=1,2,3.
\end{equation}
The potential part of the gravitational action is
given by
\begin{equation}\label{eq7}
        S_V=\int d^3xdt\,N\sqrt{\gamma}\;V[\gamma_{ij}],
\end{equation}
where $V[\gamma_{ij}]$ is a scalar function which depends on the
spatial metric and its spatial derivatives
\begin{eqnarray}
        V[\gamma_{ij}]&=&g_0\zeta^6+g_1\zeta^4R+g_2\zeta^2R^2+g_3\zeta^2R_{ij}R^{ij}\nonumber\\
        &&+g_4R^3+g_5RR_{ij}R^{ij}+g_6R^i_{\,\,j}R^j_{\,\,k}R^k_{\,\,i}\nonumber\\\label{eq8}
        &&+g_7R\nabla^2R+g_8\nabla_iR_{jk}\nabla^iR^{jk}.\hspace{1cm}
\end{eqnarray}
Here we used $\zeta$ to ensure the coupling constants $g_k$s to be
all dimensionless. The full HL action that we consider throughout
the paper is
\begin{eqnarray}
        S_{HL}&=&\frac{M_P^2}{2}\int
        d^3xdt\,N\sqrt{\gamma}(K_{ij}K^{ij}-\lambda
        K^2+R-2\Lambda\nonumber\\
        &&\quad-g_2\,M_P^{-2}\,R^2-g_3\,M_P^{-2}\,R_{ij}R^{ij}\nonumber\\
        &&\quad-g_4\,M_P^{-4}\,R^3-g_5\,M_P^{-4}\,RR_{ij}R^{ij}\nonumber\\
        &&\quad-g_6\,M_P^{-4}\,R^i_{\,\,j}R^j_{\,\,k}R^k_{\,\,i}-g_7\,M_P^{-4}\,R\nabla^2R\nonumber\\\label{eq9}
        &&\quad-g_8\,M_P^{-4}\,\nabla_iR_{jk}\nabla^iR^{jk}),
\end{eqnarray}
where we set $c=\hbar=1$ and $M_P=\frac{1}{\sqrt{8\pi G}}$
\cite{Lv,vakili-kord,HL}.

The solutions of the Einstein field equations can be classified with
respect to the type of singularities as follows: (a) Quasiregular
singularities: The observer measures no divergent physical quantity,
except when its world line arrives at the singularity, e.g., the
conical singularity of a cosmic string. (b) Scalar curvature
singularities: The observer feels diverging tidal forces when
approaching the singularity, e.g., the singularity inside a black
hole and the big bang singularity in FRW cosmology. (c) Nonscalar
singularities: The observer experiences unbounded tidal forces just
along some world-line curves, e.g., whimper cosmologies
\cite{HL,sing}. Based on the energy conditions which imply that the
gravity must be attractive, singularities are unavoidable in general
relativity. In this regard, cosmological models that contain
nonexotic fluids (radiation or dust), exhibit an initial
singularity, the so-called the big bang singularity. Since this fact
is unavoidable in general relativity, it is hoped that the quantum
theory of gravitation will solve this issue and lead to the
avoidance of the singularities in the quantum domain. In spite of
the nonexistence of the completely acceptable theory of quantum
gravity, it is shown that various approaches that combine the laws
of quantum mechanics with the general relativity could be able to
partially solve this problem \cite{HL}.

In this paper, we study the HL quantum cosmology in the presence of
the Chaplygin gas. Note that the Chaplygin gas HL classical
cosmology has been studied in Refs.~\cite{ChGHL1,ChGHL2,ChGHL3}.
Bertolami and Zarro studied the projectable HL gravity in the
context of the minisuperspace model of quantum cosmology for a FRW
Universe without matter \cite{Berto}. Also, the HL quantum cosmology
in the presence of the perfect fluid has been recently investigated
in Ref.~\cite{HL}. Here we show that the existence of the Chaplygin
gas results in a different Schr\"{o}dinger-Wheeler-DeWitt (SWD)
equation. Using the Schutz formalism, we investigate the time
evolution of the expectation value of the scale factor in the
isotropic and homogenous Universe described by the FRW spacetime. In
the framework of the Schutz formalism, the variable associated to
the degrees of freedom of matter plays the role of time and leads to
a well-define Hilbert space structure.

Notice that the proposed approach is a poor approximation of the
real dynamics of reducing the brane physics into Ho\v{r}ava-Lifshitz
gravity plus the generalized Chaplygin gas. Moreover, the HL gravity
model is not fully compatible with general relativity, but it is
discussed in order to show that a period of inflation can be
obtained. Indeed, the HL theory does not exactly recover general
relativity at low energy. However, it mimics general relativity plus
dark matter \cite{Muko}. On the other hand, the nonprojectable
Ho\v{r}ava-Lifshitz gravity is equivalent to Einstein-ether theory
with a hypersurface orthogonal ether in the IR limit (see, e.g.,
Ref.~\cite{Soti} for more detail). In Sec.~\ref{sec2}, we construct
Chaplygin gas HL quantum cosmology in minisuperspace in terms of
velocity potential variables. In Sec.~\ref{sec3}, we obtain
classical solutions and exhibit the existence of singularities in
the classical domain. Then we construct wave packets and find the
time evolution of the expectation values of the scale factor to
address the existence of singularity-free behavior of the solutions
at the quantum level. In Sec.~\ref{sec4}, we present our
conclusions.

\section{Quantum cosmology in minisuperspace}\label{sec2}
The action for the HL gravity  in the presence of Chaplygin gas and
in the Schutz formalism is given by
\begin{eqnarray}\label{eq10}
        S=S_{HL}+S_{p},
\end{eqnarray}
where $S_{HL}$ is defined in Eq.~(\ref{eq9}) and $S_p$ is the action
of the Chaplygin gas \cite{QG}:
\begin{eqnarray}\label{eq11}
        S_p=\int d^4x\sqrt{-g}\;p.
\end{eqnarray}
Since our main purpose is to study FRW cosmology, we set
$\bar{\alpha}=\beta=0$ due to the symmetry of the FRW model. So, in the
rest frame, the four-velocity of the fluid can be written as
$u_\nu=N\delta^0_\nu$ \cite{QG} which leads to
\begin{eqnarray}\label{eq12}
        \mu=\frac{\dot{\phi}+\theta\dot{S}}{N}.
\end{eqnarray}

According to Ref.~\cite{QG}, thermodynamical relations for the
Chaplygin gas are given by
\begin{eqnarray}
        \rho &=& \rho_0[1+\Pi],\nonumber\\
        \mu&=&1+\Pi+\frac{p}{\rho_0},\nonumber\\
        \tau dS &=&   d\Pi+p\,d\left(\frac{1}{\rho_0}\right),\nonumber\\\label{eq13}
        &=&\frac{(1+\Pi)^{-\alpha}}{1+\alpha}d\left[(1+\Pi)^{1+\alpha}-\frac{A}{\rho_0^{1+\alpha}
        }\right],
\end{eqnarray}
where $\tau$ and $S$ are defined as
\begin{eqnarray}\label{eq14}
  \tau &=& \frac{(1+\Pi)^{-\alpha}}{1+\alpha}, \\\label{eq15}
  S &=& (1+\Pi)^{1+\alpha}-\frac{A}{\rho_0^{1+\alpha}}.
\end{eqnarray}
Therefore, the equation of state, particle number density, and energy
density take the following forms, respectively,
\begin{eqnarray}
   p&=&-A\left[\frac{1}{A}\left(1-\frac{\,\,\mu^{\frac{\alpha+1}{\alpha}}}
{S^{1/\alpha}}\right)\right]^{\frac{\alpha}{\alpha+1}},\label{eq16}\\
  \rho &=& \left[\frac{1}{A}\left(1-\frac{\,\,\mu^{\frac{\alpha+1}{\alpha}}}{S^{1/\alpha}} \right) \right]^{\frac{-1}{\,\,1+\alpha}},\label{eq17}\\
  \rho_0 &=& \frac{\rho+p}{\mu}.\label{eq18}
\end{eqnarray}

The FRW metric is
\begin{eqnarray}\label{eq19}
        ds^2=-N(t)^2dt^2+a(t)^2\left(\frac{dr^2}{1-kr^2}+r^2d\Omega^2\right),
\end{eqnarray}
where $a(t)$ is the scale factor, $d\Omega^2$ is the metric for the
unit sphere and $k=-1,0,1$ denotes the open, flat, and closed
Universes, respectively. Thus, the three-metric is
$\gamma_{ij}=a^2\textmd{diag}\left(\frac{1}{1-kr^2},r^2,r^2\sin^2\theta\right)$,
and $\sqrt{-g}=N\sqrt{\gamma}=Na^3\frac{r^2\sin\theta}
{\sqrt{1-kr^2}}$. The Ricci curvature tensor $R_{ij}$ and the
exterior curvature tensor $K_{ij}$ are given by
\begin{eqnarray}\label{eq21}
        R_{ij}=\frac{2k}{a^2}\gamma_{ij},\qquad
        K_{ij}=\frac{\dot{a}}{Na}\gamma_{ij},
\end{eqnarray}
and the total action takes the form (in units where $16\pi G = 1$)
\begin{eqnarray}\label{eq22}
        S&=&\int dt\int d^3x\frac{r^2\sin\theta}{\sqrt{1-kr^2}}
        \Bigg\{-3(3\lambda-1)\frac{a\dot{a}^2}{N}+6Nka\nonumber\\
        &&-2N\Lambda a^3-\frac{12Nk^2}{a}\frac{3g_2+g_3}{M_P^2}
        -\frac{24Nk^3}{a^3}\frac{9g_4+3g_5+g_6}{M_P^4}\nonumber\\
        &&-Na^3A\left[\frac{1}{A}\left(1-\frac{(\dot{\phi}+\theta\dot{S})
        ^\frac{\alpha+1}{\alpha}}{N^\frac{\alpha+1}{\alpha}S^\frac{1}{\alpha}}\right)\right]
        ^\frac{\alpha}{\alpha+1}\Bigg\}.
\end{eqnarray}

Since the action does not depend on $\dot{N}$, $N$ is indeed the
Lagrange multiplier. So, it would not be surprising that the results
do not depend on how the spacetime is sliced. Now, the canonical
momenta read
\begin{eqnarray}\label{eq23}
        \begin{array}{rcl}
                p_a&=&-6(3\lambda-1)\frac{a\dot{a}}{N},\\
                p_\phi &=&a^3\left(\frac{\dot{\phi}+
                \theta\dot{S}}{NS}\right)^\frac{1}{\alpha}\left[\frac{1}{A}\left(1-\frac{(\dot{\phi}
                +\theta\dot{S})^\frac{\alpha+1}{\alpha}}
                {N^\frac{\alpha+1}{\alpha}S^\frac{1}{\alpha}}\right)\right]
                ^\frac{-1}{\alpha+1},\\
                p_{\theta}&=&0,\\
                p_S&=&\theta p_\phi,
        \end{array}
\end{eqnarray}
and the super-Hamiltonian reads \cite{QG}
\begin{eqnarray}
        \mathcal{H}&=&\frac{1}{N}\left[p_a\dot{a}+p_\phi(\dot{\phi}+\theta\dot{S})
        -\mathcal{L}\right],\nonumber\\
        &=&-\frac{p_a^2}{12(3\lambda-1)a}+g_\Lambda a^3
        -g_c(k)a+\frac{g_r(k)}{a}\nonumber\\\label{eq24}
        &&+\frac{g_s(k)}{a^3}
        +\left(Sp_\phi^{\alpha+1}+Aa^{3(\alpha+1)}\right)^\frac{1}{\alpha+1},
\end{eqnarray}
where
\begin{eqnarray}\label{eq25}
        \begin{array}{rl}
                g_\Lambda &=2\Lambda,\\
                g_c(k)&=6k,\\
                g_r(k)&=12M_P^{-2}(3g_2+g_3)k^2,\\
                g_s(k)&=24M_P^{-4}(6g_4+3g_5+g_6)k^3.
        \end{array}
\end{eqnarray}

To proceed further, let us consider the following approximation
which is valid for the early Universe \cite{BLM}:
\begin{eqnarray}\label{eq26}
        \left(Sp_\phi^{\alpha+1}+Aa^{3(\alpha+1)}\right)^\frac{1}{\alpha+1}
        &=&S^\frac{1}{\alpha+1}p_\phi\left(1+\frac{Aa^{3(\alpha+1)}}
        {Sp_\phi^{\alpha+1}}\right)^\frac{1}{\alpha+1},\nonumber\\
        &\cong&S^\frac{1}{\alpha+1}p_\phi.
\end{eqnarray}
So, the super-Hamiltonian can be written as
\begin{eqnarray}\label{eq27}
        \mathcal{H}&=&-\frac{p_a^2}{12(3\lambda-1)a}+g_\Lambda a^3
        -g_c(k)a+\frac{g_r(k)}{a}\nonumber\\
        &&+\frac{g_s(k)}{a^3}+S^\frac{1}{\alpha+1}p_\phi.
\end{eqnarray}
Now, using the canonical transformations,
\begin{eqnarray}\label{eq28}
        T&=&(\alpha+1)p_\phi^{-1}S^\frac{\alpha}{\alpha+1}p_S,\nonumber\\
        p_T&=&S^\frac{1}{\alpha+1}p_{\phi},\nonumber\\
        \{T,p_T\}&=&+1,
\end{eqnarray}
the super-Hamiltonian takes the following form:
\begin{eqnarray}\label{eq29}
        \mathcal{H}&=&-\frac{p_a^2}{12(3\lambda-1)a}+g_\Lambda a^3
        -g_c(k)a+\frac{g_r(k)}{a}\nonumber\\
        &&+\frac{g_s(k)}{a^3}+p_T.
\end{eqnarray}
Here, $p_T$ is the only remaining canonical variable associated with
the matter. The classical dynamics of the system is governed by the
Poisson brackets, namely, $\dot{A}=\{A,N\mathcal{H}\}$. In the
quantum domain, we impose the standard quantization condition on all
canonical momenta, i.e., $p_a=-i\frac{\partial}{\partial a}$ and
$p_T=-i\frac{\partial}{\partial T}$. Thus, the SWD equation is given
by
\begin{eqnarray}\label{eq30}
       \frac{\partial^2\Psi}{\partial a^2}&&+12(3\lambda-1)\Big(g_\Lambda a^4-g_c(k)a^2+g_r(k)+\frac{g_s(k)}{a^2}\Big)\Psi\nonumber\\
        &&=12i(3\lambda-1)a\frac{\partial\Psi}{\partial T}.
\end{eqnarray}
Note that the Hermicity condition for the Hamiltonian implies the
following inner product for the wave functions \cite{SWD,FRW}:
\begin{eqnarray}\label{eq31}
        (\Psi_1,\Psi_2)=\int_0^\infty da\;a\;\Psi_1^\ast(a)\Psi_2(a).
\end{eqnarray}

For the late times we have $Sp_{\phi}^{\alpha+1}\ll
Aa^{3(\alpha+1)}$ so that
\begin{eqnarray}
\hspace{-.5cm}
\left(Sp_{\phi}^{\alpha+1}+Aa^{3(\alpha+1)}\right)^\frac{1}{\alpha+1}
        \cong A^{\frac{1}{\alpha+1}}a^3
        +\frac{1}{\alpha+1}A^{-\frac{\alpha}{\alpha+1}}a^{-3\alpha}S{p_{\phi}}^{\alpha+1},\hspace{.1cm}
\end{eqnarray}
and the super-Hamiltonian takes the form
\begin{eqnarray}
\mathcal{H}&=&-\frac{p_a^2}{12(3\lambda-1)a}+g_\Lambda
a^3-g_c(k)a+\frac{g_r(k)}{a}+\frac{g_s(k)}{a^3}\nonumber\\
        &&+A^{\frac{1}{\alpha+1}}a^3+\frac{A^{-\frac{\alpha}{\alpha+1}}a^{-3\alpha}S{p_{\phi}}^{\alpha+1}}{\alpha+1}.
\end{eqnarray}
The following canonical transformations,
\begin{eqnarray}
T=-(\alpha+1)A^{\frac{\alpha}{\alpha+1}}{p_{\phi}}^{-(\alpha+1)}p_S,\quad
p_T=\frac{1}{\alpha+1}A^{-\frac{\alpha}{\alpha+1}}S{p_{\phi}}^{\alpha+1},
\end{eqnarray}
simplify the super-Hamiltonian to
\begin{eqnarray}
\mathcal{H}&=&-\frac{p_a^2}{12(3\lambda-1)a}+\left(g_\Lambda+A^{\frac{1}{\alpha+1}}\right)a^3
-g_c(k)a\nonumber\\
&&+\frac{g_r(k)}{a}+\frac{g_s(k)}{a^3}+\frac{p_T}{a^{3\alpha}}.
\end{eqnarray}
Now, imposing the standard quantization conditions, we obtain the
SWD equation as
\begin{eqnarray}
&&\frac{\partial^2\Psi}{\partial
a^2}+12(3\lambda-1)a\bigg(\left(g_\Lambda+A^{\frac{1}{\alpha+1}}\right)a^3
-g_c(k)a
\nonumber\\
&&\quad+\frac{g_r(k)}{a}+\frac{g_s(k)}{a^3}\bigg)\Psi
=12i(3\lambda-1)a^{1-3\alpha}\frac{\partial\Psi}{\partial T}.
\end{eqnarray}
Demanding that the Hamiltonian operator $\mathcal{H}$  be
self-adjoint, the inner product relation takes the form \cite{FRW}
\begin{eqnarray}
(\Phi,\Psi)=\int_0^{\infty}da\,a^{1-3\alpha}\,\Phi(a)^*\Psi(a).
\end{eqnarray}
To investigate the singularity problem, we only present the
solutions for the early Universe in Sec.~\ref{sec3}.

Moreover, the wave functions are supposed to obey the boundary
conditions
\begin{eqnarray}\label{eq32}
        \Psi(a=0,T)&=&0,\qquad\mbox{Dirichlet B.C.},\\\label{eq33}
        \left.\frac{\partial\Psi(a,T)}{\partial a}\right|_{a=0}
        &=&0,\qquad\mbox{Neumann B.C.},
\end{eqnarray}
where the first condition is called the Dewitt boundary condition to
avoid the singularity in the quantum domain.

\section{Classical and quantum mechanical solutions}\label{sec3}
The classical equations of motion are governed by
\begin{eqnarray}\label{cl1}
        \dot{a}&=&\{a,N\mathcal{H}\}=-\frac{Np_a}{6(3\lambda-1)a},\\\label{cl2}
        \dot{p}_a&=&\{p_a,N\mathcal{H}\}=N\,\left(-\frac{p_a^2}{12(3\lambda-1)a^2}
        -3g_{\Lambda}a^2\right.\nonumber\\
        &&\qquad+g_c(k)\left.+\frac{g_r(k)}{a^2}+\frac{3g_s(k)}{a^4}\right),\\\label{cl3}
        \dot{T}&=&\{T,N\mathcal{H}\}=N,\\\label{cl4}
        \dot{p}_T&=&\{p_T,N\mathcal{H}\}=0\quad\rightarrow p_T=\mathrm{const},
\end{eqnarray}
and the constraint equation $\mathcal{H}=0$.
By taking $\Psi(a,T)=e^{-iET}\psi(a)$, the time-independent SWD equation takes the form
\begin{eqnarray}\label{tiSWD}
        \frac{d^2\psi(a)}{da^2}+12(3\lambda-1)\Big(g_{\Lambda}a^4-g_c(k)a^2&&\nonumber\\
        +g_r(k)+\frac{g_s(k)}{a^2}-Ea\Big)\psi(a)&=&0.
\end{eqnarray}
Now, we present classical and quantum mechanical solutions for
various values of $g_k$s in the following subsections.

\subsection{Case $g_s(k)\neq0$}\label{seca}
In the case $g_s(k)\neq0$ where all other $g_k$s are zero, the
classical equations of motion in the gauge $N=1$ read
\begin{eqnarray}\label{acl1}
        \dot{a}&=&-\frac{p_a}{6(3\lambda-1)a},\\\label{acl2}
        \dot{p}_a&=&-\frac{p_a^2}{12(3\lambda-1)a^2}+\frac{3g_s(k)}{a^4},\\\label{acl3}
        \dot{T}&=&1.
\end{eqnarray}
After eliminating $p_a$ from Eqs.~(\ref{acl1}) and (\ref{acl2}), we
obtain
\begin{eqnarray}\label{acl5}
        T&=&t,\\\label{acl6}
        \ddot{a}+\frac{\dot{a}^2}{2a}&=&-\frac{g_s}{2(3\lambda-1)}\frac{1}{a^5}.
\end{eqnarray}
Moreover, since $\mathcal{H}=0$, we find
\begin{eqnarray}\label{acl7}
        \dot{a}^2=\frac{1}{3(3\lambda-1)}\left(\frac{g_s}{a^4}+\frac{p_T}{a}\right).
\end{eqnarray}
For $\lambda>\frac{1}{3}$ and $g_s>0$, the solution of
Eqs.~(\ref{acl6}) and (\ref{acl7}) reads
\begin{eqnarray}\label{acl9}
        a(t)=(\alpha t^2+\beta t)^{\frac{1}{3}},
\end{eqnarray}
where $\alpha=\frac{3p_T}{4(3\lambda-1)}$ and $\beta=\sqrt{\frac{3g_s}{3\lambda-1}}$.

\begin{figure}[t]
    \begin{center}
          \includegraphics[width=0.45\textwidth]{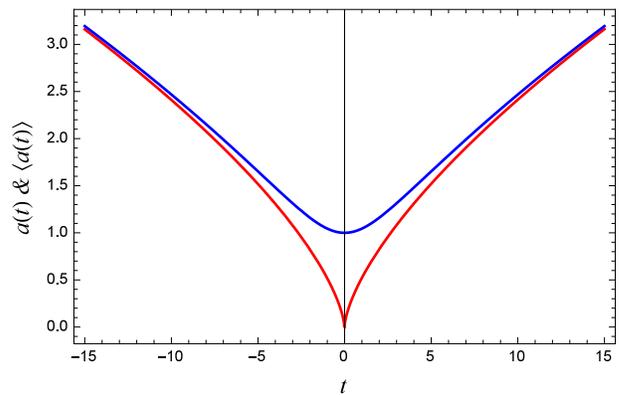}
          \caption{The behavior of $a(t)$ (red line) and $\langle a\rangle(T)$ (blue line) for $g_s(k)\neq0$ in the gauge $N=1$, i.e., $T=t$. The
          bouncing behavior of the quantum mechanical solution is apparent at $t=0$.}
          \label{fig:f1}
    \end{center}
\end{figure}

For this case, the time-independent SWD equation is
\begin{eqnarray}\label{a1}
        \frac{d^2\psi(a)}{da^2}+12(3\lambda-1)\left(-Ea+\frac{g_s(k)}{a^2}\right)\psi(a)=0.
\end{eqnarray}
So, the solutions are given by
\begin{eqnarray}\label{a2}
        \psi(a)&=&C_1\;\sqrt{a}\;
        \mathrm{J}_{\nu}\left(\frac{4i}{3}\sqrt{3(3\lambda-1)E}\,a^\frac{3}{2}\right)\nonumber\\
        &&+\;C_2\;\sqrt{a}\;\mathrm{Y}_{\nu}\left(\frac{4i}{3}\sqrt{3(3\lambda-1)E}\,a^\frac{3}{2}\right),
\end{eqnarray}
where $\mathrm{J}_{\nu}(x)$ and $\mathrm{Y}_{\nu}(x)$ are Bessel
functions with order $\nu=\frac{1}{3}\sqrt{1-48(3\lambda-1)g_s}$.
Thus, the wave function that satisfies the DeWitt boundary condition
reads
\begin{eqnarray}\label{a3}
        \Psi_{E}(a,T)=e^{-iET}\sqrt{a}\,\mathrm{J}_{\nu}\left(\frac{4i}{3}\sqrt{3(3\lambda-1)E}
        \,a^{\frac{3}{2}}\right).
\end{eqnarray}
Notice that, in the context of  general relativity and for the FRW
quantum cosmology with zero spatial curvature, namely, $k=0$, the
order of the Bessel functions is $\nu=1/3$ \cite{SWD}. Now, we
construct a wave packet with asymptotic classical behavior upon
choosing the appropriate weight function:
\begin{eqnarray}\label{a4}
        \Psi(a,T)=\int_0^{\infty}dE\,A(E)\,\Psi_{E}(a,T).
\end{eqnarray}
To this end, we use the new variable
$x=\frac{4i}{3}\sqrt{3(3\lambda-1)E}$ and choose the weight function
$A(x)=x^\nu e^{-\sigma x^2}$ where $\sigma$ is an arbitrary positive
constant. So, we have
\begin{eqnarray}\label{a5}
        \Psi(a,T)&=&-\frac{3\sqrt{a}}{8(3\lambda-1)}\nonumber\\
        &&\times\int_0^{\infty}dx\,x^{\nu+1}\,e^{-Zx^2}\,\mathrm{J}_{\nu}\left(xa^{\frac{3}{2}}\right),
\end{eqnarray}
where $Z=\sigma-i\frac{3T}{16(3\lambda-1)}$. Now, using the relation
\begin{eqnarray}
        \int_0^\infty dt\,\,t^{\nu+1}\,e^{-a^2t^2}\,\mathrm{J}_{\nu}(bt)
        =\frac{b^\nu}{(2a^2)^{\nu+1}}\,e^{-\frac{b^2}{4a^2}},
\end{eqnarray}
we find the squared integrable wave packet
\begin{eqnarray}\label{a6}
        \Psi(a,T)=-\frac{3}{8(3\lambda-1)}\frac{a^{\frac{3\nu+1}{2}}}{(2Z)^{\nu+1}}e^{-\frac{a^3}{4Z}},
\end{eqnarray}
and the expectation value of the scale factor is
given by
\begin{eqnarray}\label{a7}
        \langle a\rangle(T)=\frac{\int_0^\infty da\;a^2\;\Psi^\ast(a,T)\Psi(a,T)}
        {\int_0^\infty da\;a\;\Psi^\ast(a,T)\Psi(a,T)}.
\end{eqnarray}
Therefore, we obtain
\begin{eqnarray}\label{a8}
        \langle a\rangle(T)=\frac{\Gamma\left(\nu+\frac{4}{3}\right)}{\Gamma(\nu+1)}
        \left(2\sigma+\frac{9 T^2}{128\sigma(3\lambda-1)^2}\right)^\frac{1}{3}.
\end{eqnarray}
Since $T=t$, for
$\sigma=\frac{3}{32(3\lambda-1)p_T}\left[\frac{\Gamma\left(\nu+\frac{4}{3}\right)}{\Gamma(\nu+1)}\right]^3$,
the Universe avoids the singularity in early times and
asymptotically tends to the classical solution at late times (see
Fig.~\ref{fig:f1}). Moreover, this model predicts an accelerated
Universe at the late times. In fact, Eq.~(\ref{a8}) could be a sign
of a nonsingular behavior of the model as well as general relativity
quantum cosmology with proper initial conditions. Notice that the
nonsingular behavior is due to the Dewitt boundary condition, i.e.,
$C_2=0$, and the Gaussian smearing function is chosen to represent
the classical-quantum correspondence for large $t$. The quantum
regime of the HL theory of gravity can also provide a suitable
framework for the description of the asymptotic darkness of the
visible Universe \cite{E. Elizalde}.

Note that, various theories of quantum gravity, such as string
theory, loop quantum gravity, and black-hole physics all predict the
existence of a minimal length scale proportional to the Planck
length. The above bouncing behavior could  also be related to the
presence of a minimal length, namely, $\langle
a\rangle_{\mathrm{min}}=\langle
a\rangle(0)=\left[\frac{\Gamma\left(\nu+\frac{4}{3}\right)}{\Gamma(\nu+1)}\right]^2
\left(\frac{16(3\lambda-1)p_T}{3}\right)^{-\frac{1}{3}}$ in
agreement with various quantum gravity proposals.


\bigskip

\subsection{Case $g_r\neq0$}\label{secb}
In this case, the classical equations of motion in the gauge
$N=a(t)$ read
\begin{eqnarray}\label{airycl1}
        \dot{a}&=&-\frac{p_a}{6(3\lambda-1)},\\\label{airycl2}
        \dot{p}_a&=&-\frac{{p_a}^2}{12(3\lambda-1)a}+\frac{g_r}{a},\\\label{airycl3}
        \dot{T}&=&a(t).
\end{eqnarray}
After eliminating $p_a$ from Eqs.~(\ref{airycl1}) and
(\ref{airycl2}), we obtain
\begin{eqnarray}\label{airycl4}
        \ddot{a}-\frac{\dot{a}^2}{2a}+\frac{g_r}{6(3\lambda-1)a}=0.
\end{eqnarray}
Also, the constraint $\mathcal{H}=0$ leads to
\begin{eqnarray}\label{airycl5}
        \dot{a}^2=\frac{1}{3(3\lambda-1)}(g_r+p_T\,a).
\end{eqnarray}
For $\lambda\neq\frac{1}{3}$, the solution reads
\begin{eqnarray}\label{airycl6}
        a(t)=\alpha t^2+\beta t,
\end{eqnarray}
where $\alpha=\frac{p_T}{12(3\lambda-1)}$ and
$\beta=\sqrt{\frac{g_r}{3(3\lambda-1)}}$.

For the quantum solution, the time-independent SWD equation reads
\begin{eqnarray}\label{airy1}
        \frac{d^2\psi}{da^2}-12(3\lambda-1)E\left(a-\frac{g_r}{E}\right)\psi=0.
\end{eqnarray}
Using the variable
$x=(12(3\lambda-1)E)^{\frac{1}{3}}\left(a-\frac{g_r}{E}\right)$, the
above equation can be written as
\begin{eqnarray}\label{airy2}
        \frac{d^2\psi}{dx^2}-x\psi=0,
\end{eqnarray}
and the solutions are the Airy functions:
\begin{eqnarray}\label{airy3}
        \psi(x)&=&C_1\;\mathrm{Ai}(x)\;+\;C_2\;\mathrm{Bi}(x).
\end{eqnarray}
Airy functions $\mathrm{Ai}(x)$ [$\mathrm{Bi}(x)$] have oscillatory
behaviors for $x<0$ ($a<\frac{g_r}{E}$) and decrease (increase)
exponentially for $x>0$ ($a>\frac{g_r}{E}$) which shows that
$\mathrm{Bi}(x)$ is physically unacceptable. Therefore, we find
\begin{eqnarray}\label{airy4}
        \psi(a,T)=\mathrm{Ai}\left[(12(3\lambda-1)E)^{\frac{1}{3}}\left(a-\frac{g_r}{E}\right)\right].
\end{eqnarray}
Now, the DeWitt boundary condition implies
\begin{eqnarray}\label{airy5}
       -\left(\frac{12(3\lambda-1)}{E_n^2}\right)^{\frac{1}{3}}g_r=r_n,
\end{eqnarray}
where $r_n$ are zeros of the Airy function. Therefore, the energy spectrum reads
\begin{eqnarray}\label{airy6}
        E_n=\pm\sqrt{12(3\lambda-1)}\left(-\frac{g_r}{r_n}\right)^{\frac{3}{2}},
\end{eqnarray}
and the time-dependent solutions take the form
\begin{eqnarray}\label{airy7}
        \Psi_n(a,T)=e^{-iE_nT}\mathrm{Ai}\left[(12(3\lambda-1)E_n)^{\frac{1}{3}}a+r_n\right].
\end{eqnarray}
Notice that, for $E>0$, the Airy's function $\mathrm{Ai}(x)$
exhibits an oscillatory behavior for $x < 0$ ($a<\frac{g_r}{E}$),
whereas for $x
> 0$ ($a>\frac{g_r}{E}$), it decreases monotonically and for large $x$ becomes an exponentially damped
function. Therefore, contrary to what is usually expected,
Eq.~(\ref{airy7}) represents a classical behavior for small $a$ and
a quantum behavior for large values of the scale factor. This is in
contrast to the usual expected results. In fact, detecting quantum
gravitational effects in large Universes is noticeable which is also
observed in the FRW, Stephani, and Kaluza-Klein models in the
context of general relativity \cite{Pedram 2007,Lemos,Colistete}.
Note that for $E>0$, the solution (\ref{airy7}) is squared
integrable.

\subsection{Case $g_{\Lambda}\neq0$ and $g_s\neq0$}
In this case, the dynamics of the classical system in the gauge
$N=1$ is governed by
\begin{eqnarray}\label{Whittakercl1}
        \dot{a}&=&-\frac{p_a}{6(3\lambda-1)a},\\\label{Whittakercl2}
        \dot{p}_a&=&-\frac{{p_a}^2}{12(3\lambda-1)a^2}-3g_{\Lambda}a^2+\frac{3g_s}{a^4},\\\label{Whittakercl3}
        \dot{T}&=&1.
\end{eqnarray}
So, we obtain
\begin{eqnarray}\label{Whittakercl4}
        \ddot{a}+\frac{\dot{a}^2}{2a}+\frac{1}{2(3\lambda-1)}\left(-g_{\Lambda}a+\frac{g_s}{a^5}\right)=0.
\end{eqnarray}
Also, the constraint $\mathcal{H}=0$ leads to
\begin{eqnarray}\label{Whittakercl5}
        \dot{a}^2=\frac{1}{3(3\lambda-1)}\left(g_{\Lambda}a^2+\frac{g_s}{a^4}+\frac{p_T}{a}\right).
\end{eqnarray}
For $\lambda\neq\frac{1}{3}$ and $g_{\Lambda}\neq0$, the solution
reads
\begin{eqnarray}\label{Whittakercl6}
        a(t)=\left[\sqrt{\frac{g_s}{g_{\Lambda}}}\sinh\chi t
        +\frac{p_T}{2g_{\Lambda}}(\cosh\chi t-1)\right]^{\frac{1}{3}},
\end{eqnarray}
where $\chi=\sqrt{\frac{3g_{\Lambda}}{3\lambda-1}}$.

For the quantum solution, the time-independent SWD equation is
\begin{eqnarray}\label{Whittaker1}
        \frac{d^2\psi(a)}{da^2}+12(3\lambda-1)\Big(g_{\Lambda}(k)a^4-Ea&&\nonumber\\
        \qquad+\frac{g_s(k)}{a^2}\Big)\psi(a)&=&0.
\end{eqnarray}
For $\lambda>\frac{1}{3}$ the solution is
\begin{eqnarray}\label{Whittaker2}
        \psi(a)=C_1\;\frac{1}{a}\;\mathrm{M}_{\mu,\nu}(\xi a^3)
        \;+\;C_2\;\frac{1}{a}\;\mathrm{W}_{\mu,\nu}(\xi a^3),
\end{eqnarray}
where $\mathrm{M}_{\mu\nu}(x)$ and $\mathrm{W}_{\mu\nu}(x)$ are
Whittaker functions upon choosing
$\mu=iE\sqrt{\frac{3\lambda-1}{3g_\Lambda}}$,
$\nu=\frac{1}{6}\sqrt{1-48(3\lambda-1)g_s}$, and
$\xi=4i\sqrt{\frac{(3\lambda-1)g_{\Lambda}}{3}}$. Now the DeWitt
boundary condition is fulfilled if we set $C_1\neq0$ and $C_2=0$.

\subsection{Case $g_c\neq0$, $g_s\neq0$, and $g_r\neq0$}
The time-independent SWD equation for this case reads
\begin{eqnarray}\label{heunb1}
        \frac{d^2\psi(a)}{da^2}+12(3\lambda-1)\Big(-g_c(k)a^2+g_r(k)&&\nonumber\\
        +\frac{g_s(k)}{a^2}-Ea\Big)\psi(a)&=&0.
\end{eqnarray}
The solution for $\lambda>\frac{1}{3}$ is given by
\begin{eqnarray}\label{heunb2}
        \psi(a)&=&C_1\;e^{\kappa(g_ca^2+Ea)}a^{\frac{1+\alpha}{2}}
        \mathrm{B}_{\alpha,\beta,\gamma,\delta}(\xi a)\nonumber\\
        &&+\;C_2\;e^{\kappa(g_ca^2+Ea)}a^{\frac{1-\alpha}{2}}\;
        \mathrm{B}_{-\alpha,\beta,\gamma,\delta}(\xi a),
\end{eqnarray}
where $\mathrm{B}_{\alpha,\beta,\gamma,\delta}(x)$ is the
biconfluent Heun function \cite{heun}, and
$\kappa=\sqrt{\frac{3(3\lambda-1)}{g_c}}$,
$\alpha=\sqrt{1-48(3\lambda-1)g_s}$,
$\beta=i\left(\frac{12(3\lambda-1)}{g_c^3}E\right)^{\frac{1}{4}}E$,
$\gamma=-\sqrt{\frac{3(3\lambda-1)}{g_c^3}}\frac{4g_rg_c+E^2}{2g_c}$,
$\delta=0$, and $\xi=i(12(3\lambda-1)g_c)^\frac{1}{4}$. The above
solution with $C_1\neq0$ and $C_2\neq0$ satisfies the DeWitt
boundary condition for $0<g_s<\frac{1}{48(3\lambda-1)}$.

\subsection{Case $g_{\Lambda}\neq0$, $g_c\neq0$, and $g_r\neq0$}
In this case, the time-independent SWD equation is
\begin{eqnarray}\label{heunt1}
        \frac{d^2\psi(a)}{da^2}+12(3\lambda-1)\Big(g_{\Lambda}a^4-g_c(k)a^2&&\nonumber\\
        +g_r(k)-Ea\Big)\psi(a)&=&0,
\end{eqnarray}
and the solution for $\lambda>\frac{1}{3}$ reads
\begin{eqnarray}\label{heunt2}
        \psi(a)&=&C_1\;e^{\kappa(2g_{\Lambda}a^3-3g_ca)}\;\mathrm{T}_{\alpha,\beta,\gamma}(\xi a)\nonumber\\
        &&+\;C_2\;e^{-\kappa(2g_{\Lambda}a^3-3g_ca)}\;\mathrm{T}_{\alpha,-\beta,\gamma}(-\xi a),
\end{eqnarray}
where $\mathrm{T}_{\alpha,\beta,\gamma}(x)$ is the triconfluent Heun
function \cite{heun}, and
$\kappa=\sqrt{-\frac{3\lambda-1}{3g_{\Lambda}}}$,
$\alpha=\frac{3}{2}\frac{g_c^2-4g_{\Lambda}g_r}{g_{\Lambda}}
\left(\frac{3(3\lambda-1)^2}{2g_{\Lambda}}\right)^{\frac{1}{4}}$,
$\beta=-3E\sqrt{-\frac{3(3\lambda-1)}{g_{\Lambda}}}$,
$\gamma=g_c\left(\frac{18(3\lambda-1)}{g_{\Lambda}^2}\right)^\frac{1}{3}$,
and
$\xi=\left(-\frac{16(3\lambda-1)g_{\Lambda}}{3}\right)^{\frac{1}{6}}$.
Note that $\mathrm{T}_{\alpha,\beta,\gamma}(x)$ do not obey the
DeWitt boundary condition.

\subsection{Case $g_c(k)\neq0$ and $g_r(k)\neq0$}\label{secf}
In the particular case $g_\Lambda(k)=0$ and $g_s(k)=0$, the
classical dynamics of the system in gauge $N=a$ is governed by
\begin{eqnarray}\label{eq51}
        \dot{a}&=&-\frac{p_a}{6(3\lambda-1)},\\\label{eq52}
        \dot{p}_a&=&-\frac{p_a^2}{12(3\lambda-1)a}+g_c(k)a+\frac{g_r(k)}{a}\\\label{eq53}
        \dot{T}&=&a(t),
\end{eqnarray}
so, we have
\begin{eqnarray}\label{eq55}
        \ddot{a}-\frac{\dot{a}^2}{2a}+\frac{1}{6(3\lambda-1)}\left(g_c(k)a+\frac{g_r(k)}{a}\right)=0.
\end{eqnarray}
Also, using the Hamiltonian's constraint $\mathcal{H}=0$, we find
\begin{eqnarray}\label{eq56}
        \dot{a}^2+\frac{a}{3(3\lambda-1)}\left(g_c(k)\,a-\frac{g_r(k)}{a}-p_T\right)=0.
\end{eqnarray}
Thus, the classical solution reads
\begin{eqnarray}\label{eq57}
        a(t)=\sqrt{\frac{g_r}{g_c}}\sin\omega t+\frac{p_T}{2g_c}\left(1-\cos\omega t\right),
\end{eqnarray}
where $\omega=\sqrt{\frac{g_c}{3(3\lambda-1)}}$ is the frequency of
the oscillation.

At the quantum domain, the SWD equation is
\begin{eqnarray}\label{eq58}
        \frac{d^2\psi}{da^2}+12(3\lambda-1)\left(g_r(k)-Ea-g_c(k)a^2\right)\psi=0,
\end{eqnarray}
where its solution reads
\begin{eqnarray}\label{eq59}
        \psi_n(a)=e^{-\xi(a+E_n/2g_c)^2}\;
        \mathrm{H}_n\left(\sqrt{2\xi}(a+E_n/2g_c)\right).
\end{eqnarray}
Here, $\mathrm{H}_n(x)$ are Hermite functions,
$\xi=\sqrt{3(3\lambda-1)g_c}$, and
\begin{eqnarray}\label{eq59En}
        E_n=\pm2g_c\sqrt{\frac{2n+1}{2\xi}-\frac{g_r}{g_c}}\,,\qquad
        n=0,1,2,\ldots\,.
\end{eqnarray}
Therefore, the time-dependent squared integrable solution is
\begin{eqnarray}\label{eq60}
        \Psi_n(a,T)&=&N_n\,e^{-iE_nT}\psi_n(a)\nonumber\\
        &=&N_n\,e^{-iE_nT}e^{-\xi\left(a+E_n/2g_c\right)^2}\nonumber\\
        &&\quad\times\mathrm{H}_n\left(\sqrt{2\xi}(a+E_n/2g_c)\right),
\end{eqnarray}
where
\begin{eqnarray}
N_{n}=\sqrt{\frac{-\xi}{2^{n-1}n!{\sqrt{\pi}d_{n}}}}
\end{eqnarray}
is the normalization coefficient and
$d_n=\frac{E_n}{2g_c}\sqrt{2\xi}$. The eigenfunctions (\ref{eq60})
do not satisfy either of the two boundary conditions (\ref{eq32})
and (\ref{eq33}). Indeed, $\Psi_n(a,T)$ is similar to the stationary
quantum wormholes which is defined by Hawking. However, this model
does not represent static quantum wormholes which are ruled out by
the requirement of the self-adjointness of the Hamiltonian
\cite{AFLM}.

\begin{figure}
    \begin{center}
          \includegraphics[width=0.45\textwidth]{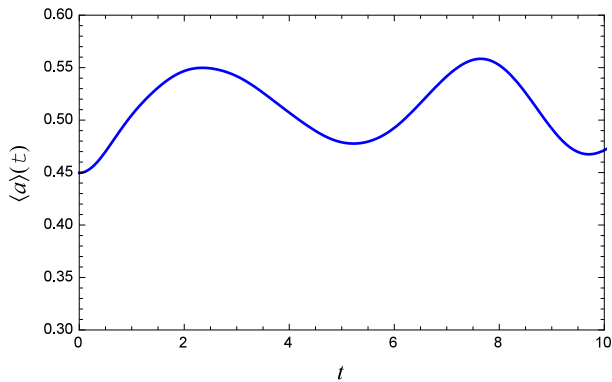}
          \caption{The evolution of the expectation value of the scale factor (\ref{time2}) with respect to the cosmic time $t$.
           We set $\lambda=1$, $g_c=2$, $g_r=0.1$, and $p_T=2g_c$.}
          \label{fig:f2}
    \end{center}
\end{figure}

Now, assuming the minus sign in Eq.~(\ref{eq59En}) due to the
condition $N_n^2>0$ and taking $\lambda>\frac{1}{3}$, $g_c\neq0$,
and $g_r<\frac{g_c}{2\xi}$, we construct wave packets upon
superimposing the above eigenfunctions (\ref{eq60}):
\begin{eqnarray}\label{eq62}
        \Psi(a,T)=\sum_{n=0}^{\infty}A_n\Psi_n(a,T).
\end{eqnarray}
Using the initial wave function,
\begin{eqnarray}\label{eq63}
        \Psi(a,T=0)\,=\,2\sqrt{2}\,\xi\,a\,e^{-\xi a^2},
\end{eqnarray}
the expansion coefficients are given by
\begin{eqnarray}
        A_n&=&\frac{N_n}{\sqrt{\xi}}\int_{-\infty}^{\infty}dx\,x^2\,e^{-\frac{x^2}{2}}
        e^{\frac{(x+d_n)^2}{2}}\mathrm{H}_n(x+d_n),\nonumber\\
        &=&\frac{N_n}{4}\sqrt{\frac{\pi}{\xi}}\,e^{-{{d_n}^2}/4}\,(d_0\,d_1)^2{d_n}^{n-2}.
\end{eqnarray}
Thus, the time evolution of the expectation value of the scale
factor, namely,
\begin{eqnarray}\label{time2}
 \langle a\rangle(T)=\frac{\int_0^\infty da\;a^2\;\Psi^\ast(a,T)\Psi(a,T)}
        {\int_0^\infty da\;a\;\Psi^\ast(a,T)\Psi(a,T)},
\end{eqnarray}
can be consequently obtained. Note that, the relation between $T$
and the cosmic time $t$ is given by
\begin{equation}\label{eq64}
        T=\frac{1}{\omega}\sqrt{\frac{g_r}{g_c}}(1-\cos\omega t)
        +\frac{p_T}{2g_c}\left(t-\frac{\sin\omega t}{\omega}\right).
\end{equation}
Figure \ref{fig:f2} shows the behavior of $\langle a\rangle(t)$ for
$\lambda=1$, $g_c=2$, $g_r=0.1$, and $p_T=2g_c$. As the figure
shows, $\langle a\rangle(t)$ exhibits an oscillatory and
nonvanishing behavior. This would be a sign that the model is
singularity free in the quantum domain.

Note that although the eigenfunctions (\ref{eq60}) do not satisfy
either of the two boundary conditions (\ref{eq32}), (\ref{eq33}),
using the initial wave function (\ref{eq63}), the expectation value
of the scale factor never tends to the singular point. This shows
that the evolution of the Universe is uniquely determined by the
initial wave packet, and no boundary condition at $a=0$ is
necessary. A similar behavior is also observed for the perfect fluid
HL quantum cosmology \cite{HL}. The oscillatory behavior of the
solution also appears in other quantum cosmological models
\cite{Monerat}.

\section{Conclusions}\label{sec4}
In this work, we have studied the Ho\v{r}ava-Lifshitz quantum
cosmology in the presence of the Chaplygin gas and in the framework
of the Schutz formalism. Due to the asymmetry of space and time in
the HL gravity, the ADM decomposition would be a natural scheme to
obtain the Hamiltonian of the model. On the other hand, the
Chaplygin gas model could describe a transition from a Universe
filled with dustlike matter to an accelerating expanding stage. As a
poor approximation the present approach reduces the brane physics
into Ho\v{r}ava-Lifshitz gravity plus the generalized Chaplygin gas.
Although the HL gravity is not fully compatible with general
relativity, we showed that a period of inflation can be achieved. In
particular, the nonprojectable Ho\v{r}ava-Lifshitz gravity is
equivalent to the Einstein-ether theory with a hypersurface
orthogonal ether at low energy \cite{Soti}.

Using Schutz variational formalism, we obtained the SWD equation
where the matter degrees of freedom played the role of time. We
exactly solved the SWD equation for various cases. For two cases,
namely, $g_s(k)\neq0$ (Sec.~\ref{seca}) and $g_c(k)\neq0$ and
$g_r(k)\neq0$ (Sec.~\ref{secf}), we constructed wave packets upon
choosing appropriate initial conditions. In both cases, unlike the
classical solutions, the expectation value of the scale factor
avoids the singularity in the quantum domain. For $g_s(k)\neq0$ the
wave function of the Universe is given by the Bessel functions with
order $\nu=\frac{1}{3}\sqrt{1-48(3\lambda-1)g_s}$, whereas the order
of the Bessel functions is $\nu=1/3$ for the Chaplygin gas FRW
quantum cosmology with zero spatial curvature ($k=0$) in the context
of general relativity. In Sec.~\ref{seca}, we constructed a wave
packet that agrees with the classical solution at late times. It is
worth mentioning that, for this case, the notion of time that arises
from Schutz formalism coincides with the cosmological time at late
times.

In Sec.~\ref{secb}, we found classical behavior for small values of
the scale factor and a quantum behavior for its large values. This
unexpected result is also observed in the FRW, Stephani, and
Kaluza-Klein models in the context of general relativity. The
bouncing behavior of the expectation value of the scale factor could
be related to the existence of a minimal length scale $\langle
a\rangle_{\mathrm{min}}=\left[\frac{\Gamma\left(\nu+\frac{4}{3}\right)}{\Gamma(\nu+1)}\right]^2
\left(\frac{16(3\lambda-1)p_T}{3}\right)^{-\frac{1}{3}}$  which is
suggested by various proposals of quantum gravity. In
Sec.~\ref{secf}, we found that the evolution of the wave function of
the Universe is uniquely determined by the initial wave packet,
i.e., no boundary condition at $a=0$ is necessary in HL quantum
cosmology.

Among various presented solutions, solutions (\ref{airycl6}), i.e.,
$a(t)=\alpha t^2+\beta t$ and (\ref{Whittakercl6}), i.e.,
$a(t)=\left[\sqrt{\frac{g_s}{g_{\Lambda}}}\sinh\chi
t+\frac{p_T}{2g_{\Lambda}}(\cosh\chi t-1)\right]^{\frac{1}{3}}$
which show the accelerated expansion of the Universe, namely,
$\ddot{a}>0$, could match the inflation scenario. In particular,
solution (\ref{airycl6}) tends to the power-law inflation
$a(t)\propto \alpha t^2$ and solution (\ref{Whittakercl6}) tends to
the exponential inflation $a(t)\propto \exp(\chi t/3)$.
Observational data of the cosmic microwave background strongly
indicate that the primordial cosmological perturbations have an
almost scale-invariant spectrum. In general relativity ($z=1$), the
scale invariance also requires inflation. Indeed, for this case, the
scale invariance is nothing but the constancy of the Hubble
expansion rate $H=\dot{a}/a$ which leads to $a\propto\exp(Ht)$. For
general $z$, the amplitude of quantum fluctuations of the scalar
field in HL gravity reads \cite{Muko}
\begin{equation}
\delta\phi\sim M\times \left(\frac{H}{M}\right)^{\frac{3-z}{2z}},
\end{equation}
where $M$ is some energy scale. For $z=3$ it is reduced to
$\delta\phi\sim M$, which means that the amplitude of quantum
fluctuations does not depend on the Hubble expansion rate. In other
words, the spectrum of cosmological perturbations in
Ho\v{r}ava-Lifshitz gravity with $z = 3$ is automatically scale
invariant even without inflation. In fact, the power law expansion
$a \propto t^p$ with $p > 1/z$ satisfies the condition. In our
study, solutions (\ref{acl9}), (\ref{airycl6}), and
(\ref{Whittakercl6}) satisfy this condition. Notice that although
solution (\ref{acl9}), i.e., $a(t)=(\alpha t^2+\beta
t)^{\frac{1}{3}}$, is not an accelerating solution, it results in
the scale-invariant spectrum in HL gravity.

\acknowledgments We would like to thank Babak Vakili and Homa
Shababi for insightful comments and suggestions. The authors are
also grateful to the referees for giving such constructive comments
which considerably improved the quality of the paper. This research
is supported by the Iran National Science Foundation (INSF), Grant
No.~93047987.


\begin{thebibliography}{99}
\bibitem{comment} Since our study is in the preinflationary phase, the current bounds on the GCG parameters are not relevant for the preinflationary phase.
\end{thebibliography}
\end{document}